# A Data-Validated Host-Parasite Model for Infectious Disease Outbreaks

Christina P. Tadiri[1*†], Jude D. Kong[2*†], Gregor F. Fussmann[1], Marilyn E. Scott[3] and Hao Wang[4*]

[1] *Department of Biology, McGill University, Montréal, QC, Canada,* [2] *Center for Discrete Mathematics and Theoretical Computer Science, Princeton, NJ, United States,* [3] *Institute of Parasitology, McGill University, Sainte-Anne-de-Bellevue, QC, Canada,* [4] *Department of Mathematical and Statistical Sciences, University of Alberta, Edmonton, AB, Canada*

The use of model experimental systems and mathematical models is important to further understanding of infectious disease dynamics and strategize disease mitigation. Gyrodactylids are helminth ectoparasites of teleost fish which have many dynamical characteristics of microparasites but offer the advantage that they can be quantified and tracked over time, allowing further insight into within-host and epidemic dynamics. In this paper, we design a model to describe host-parasite dynamics of the well-studied guppy-*Gyrodactylus* turnbulli system, using experimental data to estimate parameters and validate it. We estimate the basic reproduction number ($\mathcal{R}_0$), for this system. Sensitivity analysis reveals that parasite growth rate, and the rate at which the guppy mounts an immune response have the greatest impact on outbreak peak and timing both for initial outbreaks and on longer time scales. These findings highlight guppy population average resistance and parasite growth rate as key factors in disease control, and future work should focus on incorporating heterogeneity in host resistance into disease models and extrapolating to other host-parasite systems.

**Keywords: epidemic dynamics, mathematical model, guppy, *Gyrodactylus*, host-parasite interactions**

## INTRODUCTION

The guppy-*Gyrodactylus* system is a well-known model host-parasite system, used in numerous experimental and field studies (Scott, 1985a,b; Richards and Chubb, 1998; Cable and van Oosterhout, 2007). Guppies, *Poecilia reticulata,* are a common ovoviviparous tropical teleost fish whose abundance and ability to survive a broad range of environmental variables and availability in pet stores worldwide have made them ideal subjects for research in various disciplines. *Gyrodactylus* spp. (Monogenea) are ectoparasites which feed on the epithelial cells and mucus of many marine and freshwater teleost fish species (Bakke et al., 2007). They attach to the epidermis of their host via specialized hooks and are directly transmitted primarily by jumping to a new host during contact (Scott and Anderson, 1984; Kearn, 1994). They also reproduce directly on the host, with the developing embryo containing within itself a second developing embryo, which allows for rapid population growth of the parasite directly on an infected host (Kearn, 1994; Bakke et al., 2007). Upon infection, hosts mount an immune response, including mucus secretion (Lester, 1972), as well as a non-specific complement which kills gyrodactylids (Sato et al., 1995; Woo, 2006; Bakke et al., 2007; Robertson et al., 2017). Gyrodactylid infection can result in high rates of mortality (Van Oosterhout et al., 2003), and induce a temporary refractory period in surviving hosts (Scott and Robinson, 1984; Scott, 1985a,b).

In general, parasites are typically divided into two categories: microparasites (such as viruses and bacteria) which are microscopic and tend to proliferate and transmit rapidly, often leading to high morbidity and mortality and inducing acquired resistance in surviving hosts and therefore causing periodic epidemics, while macroparasites (such as worms and insects) often have more complex life cycles and persist in populations, often with overdispersed distributions among hosts and rarely causing severe morbidity and mortality or acquired resistance (Anderson and May, 1979; May and Anderson, 1979). Due to their rapid growth rate and infection-induced refractory period, gyrodactylids cause periodic epidemic outbreaks, making their population dynamics typical of microparasites like viruses and bacteria (Anderson and May, 1979) despite being helminths which traditionally fall into the macroparasite category. However, they have a key distinction from other typical microparasites in that parasite population size or burden is a central factor in determining host-parasite dynamics, as it directly influences transmission, mortality and several other parameters. Intrinsic population dynamics of *Gyrodactylus* sp. on isolated fish have been identified under standardized environmental conditions (Scott, 1982, 1985a,b). Both short- and long-term dynamics of *Gyrodactylus* sp. within laboratory populations of guppies have also been observed (Scott, 1985a,b; Richards and Chubb, 1998; Tadiri et al., 2013, 2016). However, the need for a comprehensive model that can describe and make predictions for this system and others like it still exists.

Traditional microparasite Susceptible, Infected, Recovered (SIR) models can effectively describe epidemic dynamics of infectious diseases for which the parasite population size is unknown or less relevant than host category of infection (Anderson and May, 1979; Grenfell and Harwood, 1997; Hagenaars et al., 2004). Yet, SIR models are less applicable to parasites such as *Gyrodactylus* spp., where parasite burden plays a crucial role in host-parasite population dynamics. Although macroparasite models directly consider parasite number, they also often overlook dynamics in parasite numbers within individual hosts (Anderson and May, 1979; Rosà et al., 2003; Cornell et al., 2004). As gyrodactylids and many other parasites do not fit neatly into the micro-/ macro-parasite dichotomy, there is a clear need for a unifying framework which considers both host and parasite populations (Gog et al., 2015). Previous efforts to mathematically describe this system using various types of models have captured basic initial epidemic dynamics, but failed to effectively describe longer-term fluctuations due to gradual loss of immunity over time (Scott and Anderson, 1984; van Oosterhout et al., 2008). Similarly, infection dynamics on individual fish have been simulated, but the broader scale transmission and population dynamics were not incorporated (van Oosterhout et al., 2008). The objective for this paper is to establish a mathematical model that effectively describes experimental data on guppy-*Gyrodactylus* dynamics, particularly with regards to host immunity waning and longer-term dynamics and to estimate the sensitivity to various parameters that we have not been able to effectively test in the laboratory. Ideally, this model can be applied to other directly transmitted, directly reproducing parasites for which parasite burden impacts host-parasite relations, with a waning immunity post-infection.

## METHODS

The guppy-*Gyrodactylus* system shares some key characteristics with common directly transmitted infectious disease dynamics (e.g., infection-induced host mortality, refractory period, infection by host-to-host contact). Therefore, we design an SIR-type model with distributed delay (which captures the varying immunity period of the guppy) to describe the dynamics of guppies and *Gyrodactylus*. Since the guppy immune response plays a crucial role in eliminating *Gyrodactylus*, we explicitly integrate the dynamics of the immune response into the model. Thereafter, the distributed delay model is converted to an equivalent system of ordinary differential equations using the linear chain approach. Next, we ensure that non-negative initial values do not give rise to a negative solution. To determine the *Gyrodactylus* basic reproduction number ($\mathcal{R}_0$), the stability analysis of the *Gyrodactylus*-free equilibrium point was performed. This threshold is particularly of use because it allows us to determine the maximum potential number of *Gyrodactylus* that will be produced due to the introduction of one *Gyrodactylus* in a *Gyrodactylus*-free population of guppies, which can help inform control measures. Model parameters unavailable in the literature were estimated by data fitting using previously published experimental data. Next, the model was validated by comparison to measurements from independent but analogous laboratory experiments. Finally, using the estimated parameters, together with the parameters from the literature, the sensitivity of the outbreak peak magnitude and the time to outbreak peak to the parameters of the model was determined. This sensitivity could be useful in determining the most influential parameters for designing control measures.

### Derivation of the Model

In this section we derive a guppy-*Gyrodactylus* interaction model with distributed delay. **Figure 1** provides a conceptual flowchart for the system. The model consists of four coupled equations tracing the rates of change of guppy population ($G$), guppy immune response ($Y$) and *Gyrodactylus* population ($X$). The guppy total population is divided into three sub-groups: susceptible ($S$), infected ($I$), and recovered ($R$) guppies. The change in number of susceptible guppies could be due to (1) birth by any guppy (we assume all guppies are born susceptible), (2) loss of immunity by a recovered guppy, (3) death of a susceptible guppy or (4) becoming infected due to contact with an infected guppy. We assume the guppy population to be homogenous and the natural birth rate of the guppy is assumed to be constant, $\alpha$. The birth rate of infected individuals is diminished by a function that is linearly proportional to the number of parasite it harbors, which we assumed to be $\eta\left(\frac{X}{I}\right) = e^{-\xi \frac{X}{I}}$, where $\xi$ is the steepness of parasite-induced fecundity reduction. Although it is unclear whether guppy fecundity is reduced by *Gyrodactylus* infection, this is the case for many infectious diseases, including those of fish (Heins et al., 2010) so we allow for it in our model, while defaulting the parameters to 0 in our simulations because in our experimental populations no birth was observed. Instead of the exponential growth assumed in typical SIR-type models, we consider a logistic growth for the guppy population because uninfected guppies exhibit density-dependent population growth up to a carrying capacity $K$ (Rose, 1959). Thus, the growth rate of the guppy is $\alpha \left(S + \eta \left(\frac{X}{I}\right) I + R\right) \left(1 - \frac{S+I+R}{K}\right)$. The rate at which susceptible fish become infected, is described by the function:

$$\beta \left(\frac{X}{I}\right) = \begin{cases} 0, & \text{if } \frac{X}{I} \leq 1 \\ b\frac{X}{I}, & \text{if } \frac{X}{I} > 1 \end{cases}$$

where $b$ is a constant. The natural death rate of guppies is assumed to be a constant $d$. The parasite mean intensity is represented as $\frac{X}{I}$. The population of infected guppies can increase when an infected guppy contacts a susceptible guppy, resulting in transmission, and decreases when any one of them dies or recovers. In addition to the natural death rate of guppies, infected guppies may also be killed by *Gyrodactylus* at a rate described by the function $\delta \left(\frac{X}{I}\right) = \mathcal{E}\frac{X}{I}$ where $\mathcal{E}$ is a constant. The recovery rate function is assumed to be directly proportional to the average immune response $Y$ (i.e., recovery function $\propto Y$) and inversely proportional to the parasite intensity $\frac{X}{I}$ (i.e., recovery function $\propto \frac{1}{X/I}$), implying that the recovery rate function is proportional to $\frac{Y}{X/I}$ (i.e., the recovery rate function is $\lambda \frac{Y}{X/I}$ where $\lambda$ is the proportionality constant). Immunity is assumed to affect both the rate of parasite population growth and the rate at which infected fish become recovered and recovered fish regain susceptibility.

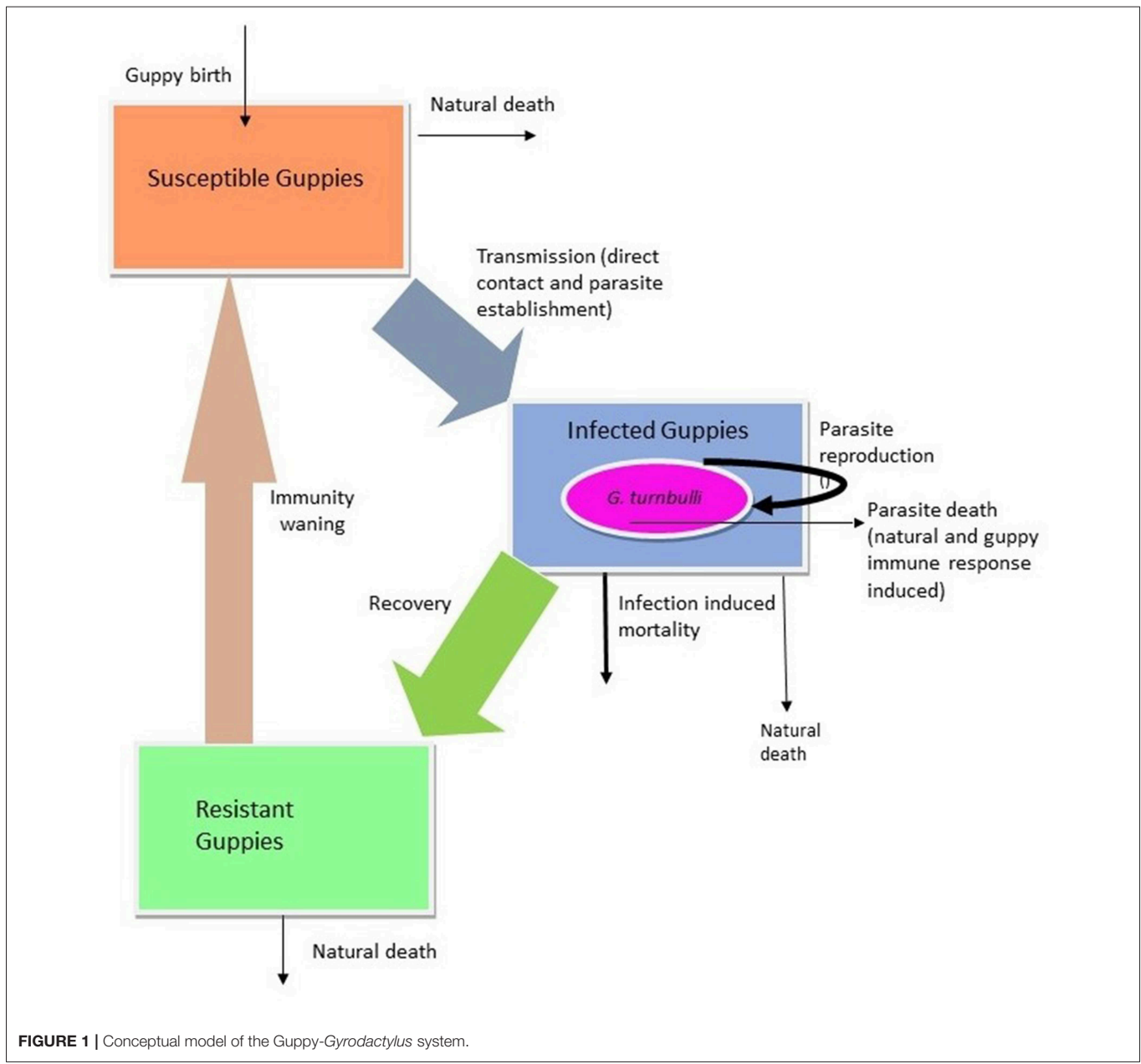

**FIGURE 1** | Conceptual model of the Guppy-*Gyrodactylus* system.

Every recovered guppy is assumed to acquire an immunity that wanes with time following initial infection. To model immunity waning, we assume that the immunity period of every fish varies from 0 to ∞ in order to capture the wide variability in the period of acquired resistance observed among many guppy populations (Scott and Robinson, 1984; Scott, 1985a,b; Richards and Chubb, 1998; Cable and van Oosterhout, 2007). We let $g(\psi)$ denote the probability density function that a fish takes exactly $\psi$ time units to lose its immunity after recovering from infection, which implies that the probability that a fish's immunity is lost $\tau$ time units after recovering from infection, is $\int_0^\tau g(\psi)\,d\psi$. Thus the probability that the fish's immunity is not lost $\tau$ time units after recovering from infection is $\int_\tau^\infty g(\psi)\,d\psi$. We assume that at a time $t - \tau$, $\lambda \frac{Y(t-\tau)I(t-\tau)}{X(t-\tau)} I(t-\tau)$ fish left the infected population compartment and joined the recovered population. The probability that these fish are still alive $\tau$ time units after leaving the infected compartment is $e^{-d\tau}$. Hence the total number of recovered fish at time t, is:

$$R(t) = \int_0^\infty \lambda \frac{Y(t-\tau)I(t-\tau)}{X(t-\tau)} I(t-\tau) e^{-d\tau} \int_\tau^\infty g(\psi)\,d\psi\,d\tau.$$

$$R(t) = \int_{-\infty}^t \lambda \frac{Y(\tau)I(\tau)}{X(\tau)} I(\tau) e^{-d(t-\tau)} \int_{t-\tau}^\infty g(\psi)\,d\psi\,d\tau. \quad (1)$$

$$\Rightarrow \frac{dR}{dt} = \lambda \frac{Y}{X} I^2 - \int_0^\infty \lambda \frac{Y(t-\tau)I(t-\tau)}{X(t-\tau)} I(t-\tau) e^{-d\tau} g(\tau)\,d\tau - dR$$

We consider $g(\tau)$ to be the density function for a gamma distribution $g(\tau) := \frac{c^n \tau^{n-1} e^{-c\tau}}{(n-1)!}$, $n = 1, 2, 3, \ldots$, $c > 0$, and $\frac{n}{c}$ to be the average duration of the immune memory. As the number of parasites increases, the guppy immune system gradually builds a defense against the parasite at a rate $f\left(\frac{X}{I}\right) = \frac{\theta \frac{X}{I}}{\kappa + \frac{X}{I}}$, proportional to the density of non-specific immune complement responsible for killing *Gyrodactylus*, where $\theta$ is the maximum rate of increase of immunity. This defense gradually reduces the guppy parasite carrying capacity. The per capita growth rate of the parasite follows a logistic growth: $\left(1 - \frac{X}{P(Y)I}\right)$, where $p(Y)$ is the average parasite carrying capacity of an infected guppy. $p(Y)$ is assumed to be an exponentially decreasing function of the average immune response of the guppy $p(Y) = re^{-\gamma Y}$, where $r$ and $\gamma$ are constants. The natural parasite death rate is assumed to be a constant, $\omega$. We assume that when a guppy dies, all the parasites on it die, since dead guppies are more likely to be predated or washed downstream (Van Oosterhout et al., 2007). In our experiments, dead fish were removed from tanks < 1 day after death in order to minimize transmission from dead fish, but it's possible some could have occurred. Thus, the total per capita parasite death rate is $\omega + d + \delta\left(\frac{X}{I}\right)$. Thus, we have the following systems of delay differential equations:

$$\frac{dS}{dt} = \alpha\left(S + \eta\left(\frac{X}{I}\right)I + R\right)\left(1 - \frac{S+I+R}{K}\right) - \frac{\beta\left(\frac{X}{I}\right)SI}{S+I+R} - dS$$
$$+ \int_0^\infty \lambda \frac{Y(t-\tau)I(t-\tau)}{X(t-\tau)} I(t-\tau) e^{-d\tau} g(\tau)\,d\tau$$

$$\frac{dI}{dt} = \frac{\beta\left(\frac{X}{I}\right)SI}{S+I+R} - \left(d + \delta\left(\frac{X}{I}\right)\right)I - \frac{\lambda YI}{X}I$$

$$\frac{dR}{dt} = \frac{\lambda YI}{X}I - \int_0^\infty \lambda \frac{Y(t-\tau)I(t-\tau)}{X(t-\tau)} I(t-\tau)e^{-d\tau} g(\tau)\,d\tau - dR$$

$$\frac{dX}{dt} = \mu X\left(1 - \frac{X}{p(Y)I}\right) - \left(d + \delta\left(\frac{X}{I}\right)\right)X - \omega X \quad (2)$$

$$\frac{dY}{dt} = Y\left(\frac{\theta \frac{X}{I}}{\kappa + \frac{X}{I}}I - \nu\right)$$

$$(S(s), I(s), R(s), X(s), Y(s))$$
$$= (\phi_1(s), \phi_2(s), \phi_3(s), \phi_4(s), \phi_5(s)), \, s \in (-\infty, 0]$$

where $\phi_i, i = 1, 2, 3, 4$ and 5 are bounded continuous functions, $\mu$ is the maximum per capita parasite growth rate. We assume that.

$$X(I < 1) = 0$$

## Reduction to an Ordinary Differential Equation Model

We assume that $g(\tau) = ce^{-c\tau}$ (i.e. $n = 1$) and apply the chain trick method (Kuang, 1993) to convert System (2) to a system of ordinary differential equations: Let

$$\overline{S} = \int_0^\infty \frac{Y(t-\tau)I(t-\tau)}{X(t-\tau)} I(t-\tau) e^{-d\tau} g(\tau)\,d\tau$$

$$= \int_0^\infty \frac{Y(t-\tau)I(t-\tau)}{X(t-\tau)} I(t-\tau) e^{-d\tau} ce^{-c\tau}\,d\tau$$

$$= \int_{-\infty}^0 \frac{Y(u)I(u)}{X(u)} I(u) e^{-d(t-u)} ce^{-c(t-u)}\,du$$

$$= ce^{-(c+d)t} \int_{-\infty}^0 \frac{Y(u)I(u)}{X(u)} I(u) e^{(d+c)u}$$

$$\Rightarrow \frac{d\overline{S}}{dt} = c\left(-(c+d)\right)e^{-(c+d)t}\int_{-\infty}^0 \frac{Y(u)I(u)}{X(u)} I(u) e^{-(d+c)u}\,du$$
$$+ ce^{-(c+d)t}e^{(c+d)t}\frac{Y(t)I(t)}{X(t)}I(t)$$
$$= -(c+d)\overline{S} + c\frac{Y(t)I(t)}{X(t)}I(t)$$

Substituting this in System (2) reduces it to the following system of ODE:

$$\frac{dS}{dt} = \alpha\left(S + \eta\left(\frac{X}{I}\right)I + R\right)\left(1 - \frac{S+I+R}{K}\right)$$
$$- \frac{\beta\left(\frac{X}{I}\right)SI}{S+I+R} - dS + \lambda\overline{S}$$

$$\frac{dI}{dt} = \frac{\beta\left(\frac{X}{I}\right)SI}{S+I+R} - \left(d + \delta\left(\frac{X}{I}\right)\right)I - \frac{\lambda YI}{X}I$$

$$\frac{dR}{dt} = \frac{\lambda YI}{X}I - \lambda\overline{S} - dR$$

$$\frac{dX}{dt} = \mu X\left(1 - \frac{X}{p(Y)I}\right) - \left(d + \delta\left(\frac{X}{I}\right)\right)X - \omega X \quad (3)$$

$$\frac{dY}{dt} = Y\left(\frac{\frac{\theta X}{I}}{\kappa + \frac{X}{I}}I - \nu\right)$$

$$\frac{d\bar{S}}{dt} = -(c+d)\bar{S} + c\frac{Y(t)I(t)}{X(t)}I(t)$$

$$(S(0), I(0), R(0), X(0), Y(0))$$
$$= (\phi_1(0), \phi_2(0), \phi_3(0), \phi_4(0), \phi_5(0))$$

$$\bar{S}(0) = \int_0^\infty \frac{\phi_4(-\tau)\phi_2(-\tau)}{\phi_3(-\tau)}\phi_2(-\tau)ce^{-(c+d)\tau}d\tau$$

## Positivity and Basic Reproduction Number

In this section, we show that non-negative initial data give rise to non-negative solutions, establish conditions for the existence and stability of the *Gyrodactylus*-free equilibrium point of the system and determine the basic reproduction number.

### Positivity

Positivity and boundedness of a model guarantee that the model is biologically well-behaved. For positivity of the System (3), we have the following theorem:

**Theorem 1.** *All solutions of System (3) are positive for all t in* $(0, \infty)$

*Proof.* We need to show that $S(t) \geq 0$, $I(t) \geq 0$, $R(t) \geq 0$, $Y(t) \geq 0$, $X(t) \geq 0$, $\bar{S}(t) \geq 0$ for $S(0) \geq 0$, $I(0) \geq 0$, $R(0) \geq 0$, $Y(0) \geq 0$, $X(0) \geq 0$ and $\bar{S}(0) \geq 0$. Note that $\frac{X}{I} = 0$, $\frac{I}{X} = 0$ and $\frac{Y}{X} = 0$ for $X < 1$ or $I < 1$. We start by proving that if $I(0) \geq 0, \Rightarrow I(t) \geq 0$ for all $t > 0$. From the second equation of Systems (3), we see that $\dot{I}(I = 0) = 0$. Thus $I(t) \geq 0$ for $t \geq 0$. Also, from the third equation, we have that $\dot{X}(X = 0) = 0$ for $X(0) \geq 0$. Hence $X(t) \geq 0$ for $t \geq 0$. From the last equation, we have that $\dot{\bar{S}}(\bar{S} = 0) = \frac{cY}{X}I^2$. Since $I(t) \geq 0$ and $\frac{X}{Y} \geq 0$ for $t \geq 0$, $I(0) \geq 0$ and $\frac{X(0)}{Y(0)} \geq 0$, we have that $\dot{\bar{S}}(\bar{S} = 0) \geq 0$, $t \geq 0$ and $\bar{S}(0) \geq 0$. The non-negativity of $R(t)$ for $t \geq 0$ follows from the integral representation (1) and non-negativity of $\frac{Y(t)}{X(t)}$. From the first equation, we have that

$$\dot{S}(S = 0) = \alpha\left(\eta\left(\frac{X}{I}\right)I + R\right)\left(1 - \frac{I + R}{K}\right) + \lambda\bar{S}.$$

From the non-negativity of $\frac{X}{I}, I, R$, we have that if $S(0) \geq 0$ implies that $S(t) \geq 0$ for $t \geq 0$.

### *Gyrodactylus*-Free Equilibrium Point (GFE) and $\mathcal{R}_0$

For the GFE we have that $X = 0$, implying that $I, R, Y$ and $\bar{S}$ are null. Plugging this in System (3), we have that $S = \frac{k(\alpha-d)}{\alpha}$. Since $S > 0$, we have that the GFE exist iff $\alpha > d$ and is given by $\left(\frac{k(\alpha-d)}{\alpha}, 0, 0, 0, 0\right)$. The linearized system corresponding to this equilibrium point is:

$$\begin{pmatrix}\dot{S}\\\dot{I}\\\dot{R}\\\dot{X}\\\dot{Y}\\\dot{\bar{S}}\end{pmatrix} = \begin{pmatrix} d-\alpha & 2d-\alpha & 2d-\alpha & d\eta-b & 0 & \lambda \\ 0 & -d & 0 & b-\varepsilon & 0 & 0 \\ 0 & 0 & -d & 0 & 0 & -\lambda \\ 0 & 0 & 0 & \mu-d-\omega & 0 & 0 \\ 0 & 0 & 0 & 0 & -\nu & 0 \\ 0 & 0 & 0 & 0 & 0 & -(c+d) \end{pmatrix} \begin{pmatrix}S\\I\\R\\X\\Y\\\bar{S}\end{pmatrix}$$

The corresponding eigenvalues are:

$$\lambda_1 = d - \alpha, \lambda_2 = -d, \lambda_3 = -d, \lambda_4 = \mu - d - \omega,$$
$$\lambda_5 = -(c+d).$$

$\lambda_1$, $\lambda_2$, $\lambda_3$ and $\lambda_5$ are all less than zero and $\lambda_4$ is less than zero iff $\mu < d + \omega$. This leads to the definition of the *Gyrodactylus* basic reproduction number, $\mathcal{R}_0 := \frac{\mu}{d+\omega}$. Observe that GFE is locally asymptotically stable iff $\mu < d + \omega$, i.e., GFE is locally asymptotically stable iff $\mathcal{R}_0 < 1$ and unstable if $\mathcal{R}_0 > 1$. Therefore, if $\mathcal{R}_0 < 1$, the parasite dies out and if $\mathcal{R}_0 > 1$, the parasite will invade the guppy population. $\mathcal{R}_0 = 1$, is a threshold below which the *Gyrodactylus* dies out and above which there is an outbreak. $\mathcal{R}_0$ has an intuitive biological interpretation: it is the average number of *Gyrodactylus* resulting from the introduction of a single *Gyrodactylus* into an otherwise *Gyrodactylus*-free population over the course of its life span.

## Parameter Estimation and Model Validation Using Independent Measurements

We used data obtained from separate laboratory previously published experiments to, respectively, estimate the model parameters not available in the literature and to test the fit of the model. To estimate the model parameters we used experimental data averaged from four groups of eight male fish where one fish per group was infected with two parasites and the infection was allowed to spread naturally throughout the tank (Tadiri et al., 2018). These fish were bred in the lab from pet-store "feeder" guppies. To test the model, we used experimental data averaged from four groups of eight fish (four males and four females) where parasites were introduced to each group via a donor juvenile fish infected with three parasites that was removed once at least three parasites had naturally transferred to the experimental fish (Tadiri et al., 2016). These fish were third-generation lab reared fish bred from 33 originally family lines originally obtained from wild population in Trinidad but mixed haphazardly in experimental tanks. In both cases, 2-3 parasites were introduced in order to keep the introduction as close to one as possible while minimizing the probability of accidental parasite death or that an old or male parasite would be introduced preventing reproduction. No difference in host-parasite dynamics was found among all-male, all-female and mixed sex groups of eight fish (Tadiri et al., 2016).

In both experiments, each fish was individually marked, and the number of parasites on each fish was counted every other day to obtain SIR numbers and total parasite population size. In our all experimental groups, no birth was observed within the 42 days. Hence, we ignore vital dynamics for guppies in the model.

To estimate the parameter values of System (3), we use the non-linear regression function nlinfit(.) in MATLAB. The function nlinfit(.) uses the Levenberg-Marquardt algorithm (Moré, 1978) to fit the solution of the biodegradation module to the data. Some of parameters used in solving System (3) namely $\omega$, $d$, and $K$, were taken from the literature (Rose, 1959; Scott and Anderson, 1984): the units, values and source of these parameters are provided in **Table 1**.

TABLE 1 | Estimated parameter values used to train and test the model and initial values from experiments.

| Parameter | Sym. | Estimate | Unit | Source |
|---|---|---|---|---|
| Initial number of susceptible guppies | S (0) | 7 | - | This study |
| Initial number of infected guppies | I (0) | 1 | - | This study |
| Initial number of recovered guppies | R (0) | 0 | - | This study |
| Initial number of parasites | X (0) | 2 | - | This study |
| Initial number of immune cells | Y(0) | 5.1440 | - | This study |
| Transmission rate | $\beta$ | 0.0468 | /day/host/parasite | This study |
| Recovery rate | $\lambda$ | 0.0080 | /day/host | This study |
| Half-saturation constant of per-capita parasite killing rate | $r$ | 176.7035 | - | This study |
| Maximum parasite killing rate | $\gamma$ | 0.1084 | /day/no immune cells | This study |
| Parasite increase rate | $\mu$ | 0.6395 | /day | This study |
| Maximum rate of immunity increase | $\theta$ | 0.0562 | /day | This study |
| Half-saturation constant of immunity increase | $\kappa$ | 7.1411 | - | This study |
| Rate of decay of immunity in the absence of parasites | $\nu$ | 0.0321 | /day | this study |
| Guppy carrying capacity | $K$ | 9.600 | /liter | Rose, 1959 |
| Steepness of distribution kernel | $c$ | 0.0100 | /day | This study |
| Parasite-induced mortality rate | $\varepsilon$ | 0.0012 | /day | This study |
| Guppy birth rate | $\alpha$ | 0.0 | /fish/day | This study |
| Natural guppy mortality rate | $d$ | 0.0049 | /day | Scott and Anderson, 1984; Kearn, 1994 |
| Natural parasite mortality rate | $w$ | 0.24 | /day | Scott and Anderson, 1984 |

*Volume of experimental tanks was 6L. As no guppy birth was observed in our experiments, we estimate α to be 0 and neglect the parasite induced fecundity reduction.*

The validity of our model in predicting *Gyrodactylus* outbreak was evaluated by using the estimated parameters in the model to generate S, I, R *Gyrodactylus* and immune response data then comparing the predicted data to measured data using the goodnessOfFit(.) function in MATLAB.

## Sensitivity Analysis

The objective of this subsection is to discuss the sensitivity of the magnitude of the initial *Gyrodactylus* outbreak peak, and time to the initial outbreak peak to the parameters of the system. For this analysis, we use the normalized forward sensitivity index (Chitnis et al., 2008):

$$\text{sensitivity index (S.I.)} = \left(\frac{\partial F^*}{\partial (parameter)}\right)\left(\frac{parameter}{F^*}\right) \quad (4)$$

where $F^*$ is the quantity being considered.

Since we do not have the explicit formula for the initial outbreak peak, or time to peak, we use central difference approximation to estimate them:

$$\frac{\partial F^*}{\partial parameter} = \frac{F^*(parameter + h) - F^*(parameter - h)}{2h} + O(h^2).$$

Letting $h = 1\%$ of the parameter value (P), Equation (4) becomes:

$$S.I. = \frac{F^*(1.01P) - F^*(0.99P)}{0.02(F*(P))} \quad (5)$$

## Longer-Term Dynamics and Generic Sensitivity Analysis

In this section, we simulate the longer-term parasite dynamics in a system that allows for guppy vital dynamics. We use an average birth rate estimated from literature of 0.4/fish/day (Rose, 1959) with no infection-induced reduction in fecundity. We equally assess the sensitivity of the generic outbreak peaks and period to the parameters of the system from the long-term simulation.

## RESULTS

### System Basic Reproduction Number and Gyrodactylus-Free Equilibrium Point

Using the parameters in **Table 1** we have that $\mathcal{R}_0 = 2.63$. Since $\mathcal{R}_0 > 1$ for the estimated parameters values, the GFE is not asymptotically stable, meaning that an introduction of one *Gyrodactylus* into a naïve guppy population would result in an outbreak. Next, we illustrate the dynamics of the system for $\mathcal{R}_0 < 1$ and for $\mathcal{R}_0 > 1$ using values very close to one. Rearranging the fourth equation of System (2), we have:

$$\frac{dX}{dt} = \mu X\left(1 - \frac{X}{p(Y)I}\right) - \left(d + \delta\left(\frac{X}{I}\right)\right)X - \omega X$$

$$= (d + \omega)(\mathcal{R}_0 - 1)X - X\left(\frac{\mu X}{p(Y)I} + \delta\left(\frac{X}{I}\right)\right) \quad (6)$$

**Figure 2** shows the long-term behavior of the *Gyrodactylus* population for $\mathcal{R}_0 = 0.9 (< 1)$ (A), $\mathcal{R}_0 = 1.1 (>$

1) (B) and $\mathcal{R}_0 = 2.63 (> 1)$ (C) respectively. When $\mathcal{R}_0 < 1$, the system will stabilize to its *Gyrodactylus* free equilibrium $\left(\frac{k(\alpha-d)}{\alpha}, 0, 0, 0, 0, 0\right)$ (Panel A). The number of *Gyrodactylus*, the number of infected guppies, the number of recovered guppies and the guppy immune compliment density tend to zero as $t$ increases. When $\mathcal{R}_0 > 1$, there will be a *Gyrodactylus* outbreak (Panels B and C). These outcomes are robust for large sets of initial values and parameter values.

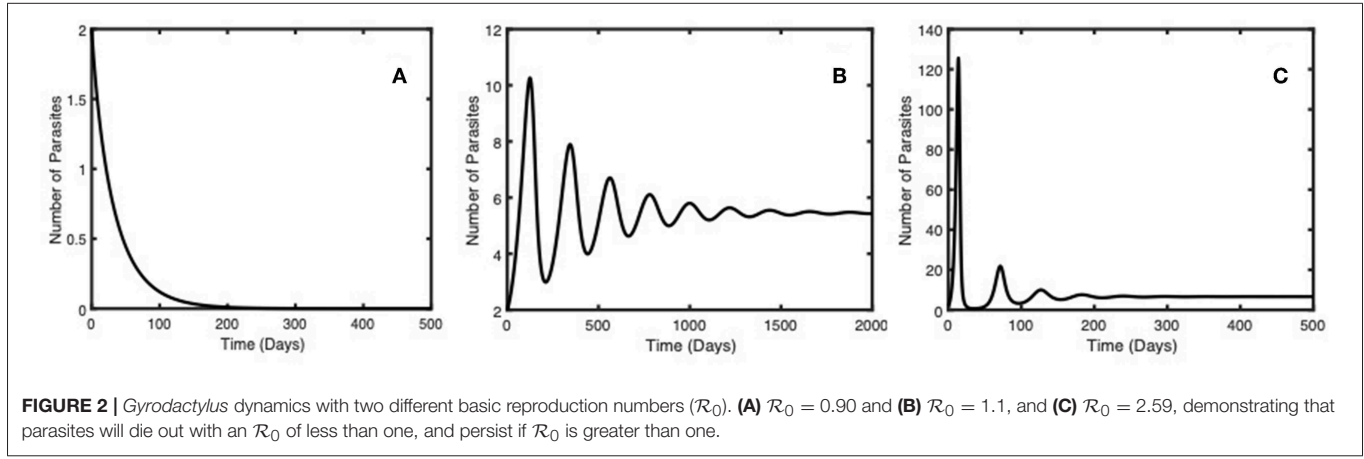

**FIGURE 2** | *Gyrodactylus* dynamics with two different basic reproduction numbers ($\mathcal{R}_0$). **(A)** $\mathcal{R}_0 = 0.90$ and **(B)** $\mathcal{R}_0 = 1.1$, and **(C)** $\mathcal{R}_0 = 2.59$, demonstrating that parasites will die out with an $\mathcal{R}_0$ of less than one, and persist if $\mathcal{R}_0$ is greater than one.

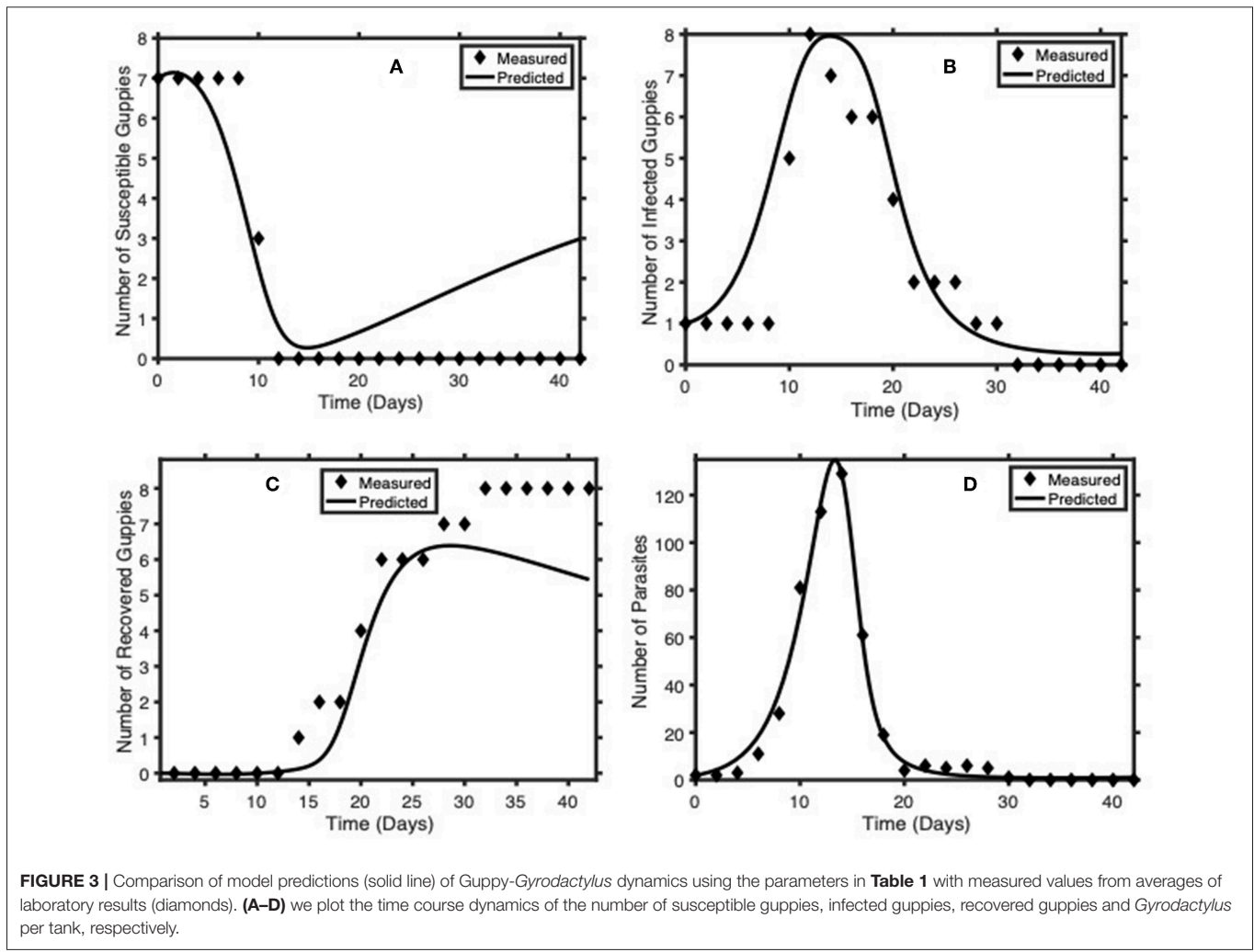

**FIGURE 3** | Comparison of model predictions (solid line) of Guppy-*Gyrodactylus* dynamics using the parameters in **Table 1** with measured values from averages of laboratory results (diamonds). **(A–D)** we plot the time course dynamics of the number of susceptible guppies, infected guppies, recovered guppies and *Gyrodactylus* per tank, respectively.

## Fitting the Model to Data

**Table 1** contains the value of the parameters obtained from fitting System (2) to the experimental data described above. **Figure 3** shows the simulated susceptible, infected, recovered guppies and *Gyrodactylus* dynamics along with measured data. We obtained a goodness-of-fit statistic (NMSE) value of 0.99. This statistic indicates that the model is able to predict the training data accurately.

## Model Evaluation Using NMSE

Using the parameter values in **Table 1**, with the procedure described above, we assess the validity of our model in predicting guppy-*Gyrodactylus* dynamics data. **Figure 4** shows a comparison between our simulated and measured data. The goodness-of-fit statistics suggests that System (2) with the given parameter values is a good fit for guppy-*Gyrodactylus* dynamics data (NMSE = 0.70).

## Sensitivity of the Magnitude of the Initial Outbreak Peak

The sensitivity indices of the magnitude of the peak of the initial outbreak measure how the magnitude of the peak of the initial outbreak depends on different parameters. **Table 2** contains the sensitivity indices of the amplitude of the first outbreak peak obtained using Equation (5). The two parameters with the greatest independent influence on the system were parasite increase rate ($\mu$), and maximum rate of increase of immunity ($\theta$).

## Sensitivity of the Time to Initial Outbreak Peak

Sensitivity indices of the time to initial outbreak peak measure how the first epidemic outbreak time depends on different parameters as seen in the **Table 3**. Similar to the sensitivity of the magnitude of the peak of the first outbreak, the parasite increase rate ($\mu$) and the maximum rate of the increase of immunity ($\theta$) were the most influential parameters.

## Longer-Term Dynamics

By allowing for natural guppy birth in our systems, we are able to simulate steady state oscillating dynamics such as those expected of epidemic infectious diseases with generic peaks and periods (**Figure 5**).

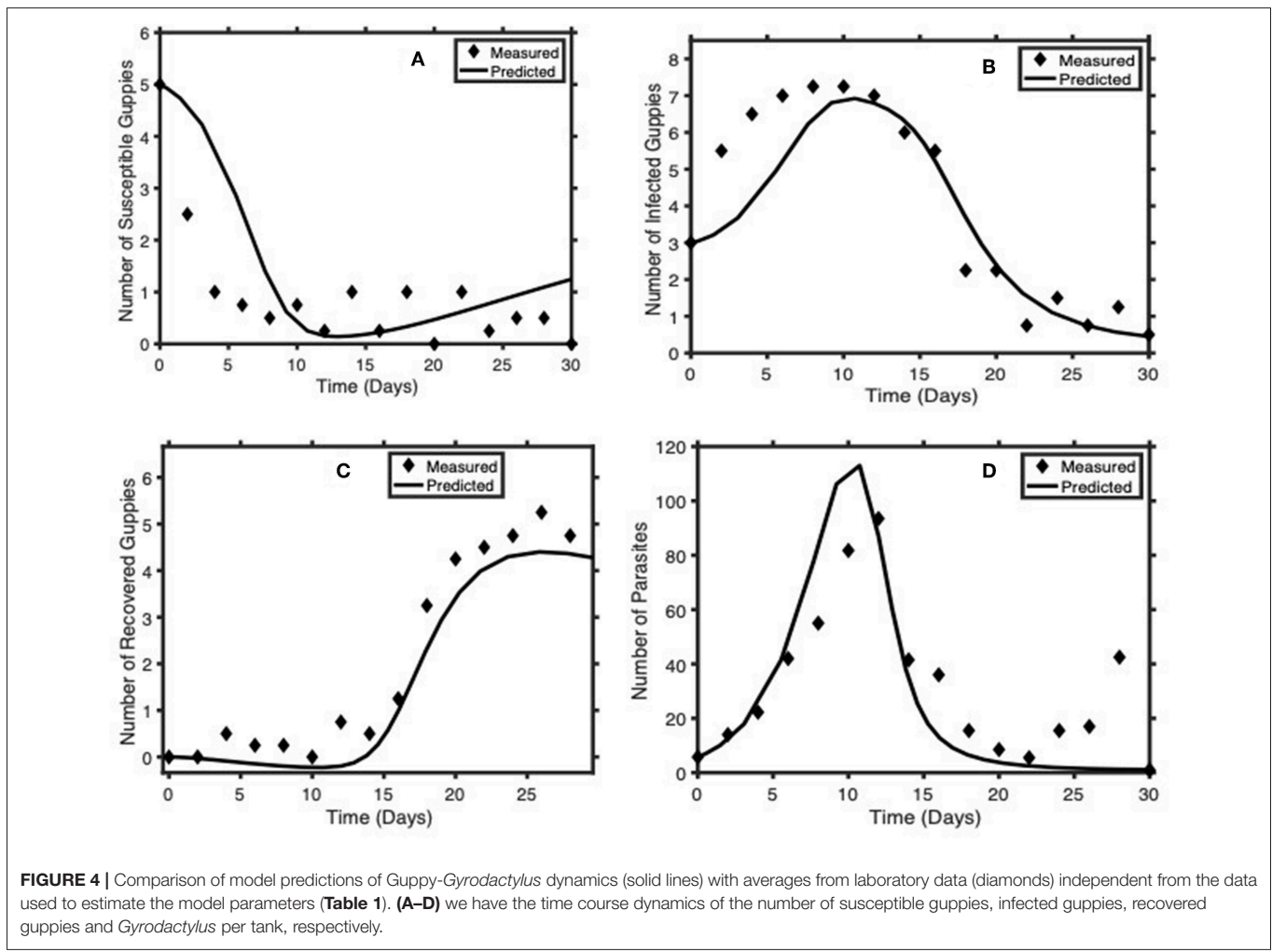

**FIGURE 4** | Comparison of model predictions of Guppy-*Gyrodactylus* dynamics (solid lines) with averages from laboratory data (diamonds) independent from the data used to estimate the model parameters (**Table 1**). **(A–D)** we have the time course dynamics of the number of susceptible guppies, infected guppies, recovered guppies and *Gyrodactylus* per tank, respectively.

TABLE 2 | The sensitivity of the magnitude of the peak of the initial outbreak to the parameters.

| Parameter | Definition | Sensitivity index |
|---|---|---|
| β | Transmission rate | 0.0130 |
| λ | Recovery rate | −0.0516 |
| μ | Parasite increase rate | 1.5662 |
| r | Half-saturation constant of per-capita parasite killing rate | 0.4707 |
| γ | Maximum parasite killing rate | −0.6369 |
| θ | Maximum rate of increase of immunity | −0.8186 |
| κ | Half-saturation constant of increase of immunity | 0.3367 |
| ν | Rate of decay of immunity in the absence of parasites | 0.2253 |
| α | Birth rate | 0.0216 |
| d | Natural guppy mortality | −0.0178 |
| K | Half-saturation constant for guppy growth | 0.2541 |
| c | 1/c is the average duration of immune memory | −0.0016 |
| ε | Parasite-induced mortality rate | −0.0568 |
| ω | Natural parasite mortality | −0.6862 |

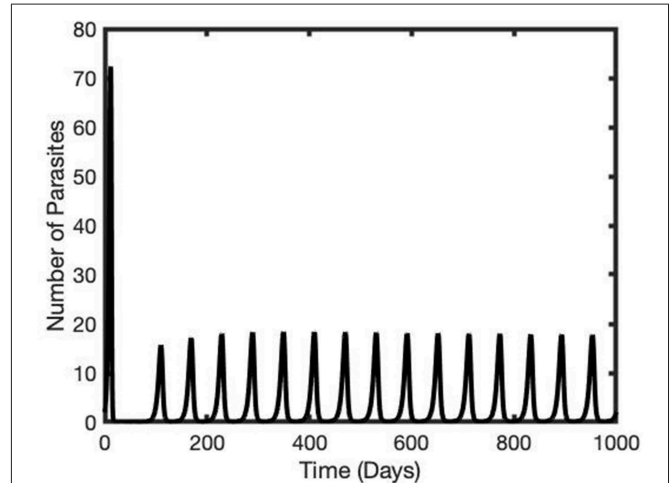

FIGURE 5 | Long-term *Gyrodactylus* dynamics when guppy birth is included in the system. The Figure was generated using the parameter values in Table 1 with α = 0.4/fish/day.

TABLE 3 | The sensitivity of the time to the initial outbreak to the parameters.

| Parameter | Definition | Sensitivity index |
|---|---|---|
| β | Transmission rate | −0.7344 |
| λ | Recovery rate | −0.1732 |
| μ | Parasite increase rate | −1.0673 |
| r | Half-saturation constant of per-capita parasite killing rate | −0.1847 |
| γ | Maximum parasite killing rate | 0.1212 |
| θ | Maximum rate of increase of immunity | −1.1184 |
| κ | Half-saturation constant of increase of immunity | −0.0880 |
| ν | Rate of decay of immunity in the absence of parasites | −0.1329 |
| α | Birth rate | 0.0593 |
| d | Natural guppy mortality | 0.0088 |
| K | Half-saturation constant for guppy growth | −0.8757 |
| c | 1/c is the average duration of immune memory | −0.0055 |
| ε | Parasite-induced mortality rate | 0.0057 |
| ω | Natural parasite mortality | −0.2868 |

TABLE 4 | The sensitivity of the magnitude of the peak of a generic outbreak to the parameters.

| Parameter | Definition | Sensitivity index |
|---|---|---|
| β | Transmission rate | −0.8688 |
| λ | Recovery rate | 0.1385 |
| μ | Parasite increase rate | 0.5096 |
| r | Half-saturation constant of per-capita parasite killing rate | 0.1465 |
| γ | Maximum parasite killing rate | −0.4790 |
| θ | Maximum rate of increase of immunity | −1.5751 |
| κ | Half-saturation constant of increase of immunity | 0.1726 |
| ν | Rate of decay of immunity in the absence of parasites | 1.9784 |
| α | Birth rate | −0.0344 |
| d | Natural guppy mortality | −0.4309 |
| K | Half-saturation constant for guppy growth | −0.5502 |
| c | 1/c is the average duration of immune memory | −0.3137 |
| ε | Parasite-induced mortality rate | −0.0754 |
| ω | Natural parasite mortality | −0.4031 |

## Sensitivity of the Generic Outbreak Peak Magnitude

Sensitivity indices of the generic outbreak peak measure how the magnitude of subsequent outbreak peaks under steady state oscillating dynamics depend on different parameters (**Table 4**). The guppy immune related parameters, $\theta$ (maximum rate of increase of immunity and $\nu$ (rate of decay of immunity) have the strongest relationship to the outbreak peak. The negative value of the sensitivity of the outbreak to $\theta$, indicates that a low value of $\theta$ will lead to a more severe parasite outbreak. The positive value of the sensitive of the outbreak peak to $\nu$, on the other hand, tells us that if the guppy immunity wanes faster, there will be a severe outbreak.

## Sensitivity of the Generic Outbreak Period

Sensitivity indices of the generic outbreak period measure how the time to subsequent outbreak peaks under steady state oscillating dynamics depend on different parameters (**Table 5**). The parasite increase rate ($\mu$) and the maximum rate of increase of immunity ($\theta$) have the strongest relationship to the outbreak period.

# DISCUSSION

In this paper, we define a mathematical model that effectively describes guppy-*Gyrodactylus* dynamics in small populations. In estimating parameters based on both literature and our own experimental data we determined our model to accurately describe the dynamics of this system. Additionally, we validated our model using a neutral data set from a separate experiment and found that it fit reasonably well. We also find the model to be mathematically and biologically sound through our analysis of $\mathcal{R}_0$, which indicates that an outbreak will occur when the $\mathcal{R}_0$ is greater than one and that the system will stabilize to a *Gyrodactylus*-free equilibrium when $\mathcal{R}_0$ is less than one. With our parameters, $\mathcal{R}_0$ was greater than one, indicating that an outbreak will occur in our system with the introduction of one parasite. This model builds of previous efforts to model this system (Scott and Anderson, 1984), but incorporates a more realistic representation of immunity. Firstly, and most importantly, we describe fish immune response to infection in its own equation, rather than assuming a linear constant to represent immunity. We specifically also describe the waning of immunity post-infection using a distributed delay function, which allows for repeated, dampening cycles of outbreaks without constant immigration of naive hosts. We also allow for host population growth rather than fixed immigration and consider a parasite-induced reduction in fecundity (Perez-Jvostov et al., 2012), which had not been investigated at the time of previous models. Longer-term simulations using population growth estimates from literature with our other parameter estimates from experiments demonstrate that our model is capable of describing oscillating parasite dynamics typical of those observed in the wild (Van Oosterhout et al., 2007). These developments are important to more accurately explaining guppy-*Gyrodactylus* dynamics and could have a broader applicability to other systems as well.

*Gyrodactylus* are a large genus of over 400 ectoparasites infecting at least 20 orders of teleost fish (Bakke et al., 2002), and our model can most directly be applied to other species of this genus. Gyrodactylids have had significant economic impact, causing epizootics in many resources fish such as carp, trout and African catfish (Woo, 2006) and, most notably Atlantic salmon fisheries, particularly in Norway in the 1960's and '70's which saw large declines due to *G. salaris* (Johnsen, 1978; Bakke et al., 2007), and efforts to recover these populations and prevent disease spread to other watersheds are still ongoing (Denholm et al., 2016). Since the basic life cycle of gyrodactylids and their relationships with their hosts are similar (Bakke et al., 2002, 2007) this model would only need reparameterization to be applied to a range of other aquaculture species. Beyond other gyrodactylids, many infectious diseases also confer immunity that decays over time. Our methods of applying a distributed delay to describe waning guppy immunity to *Gyrodactylus* are novel to this system, and can also be used for other infectious diseases with declining immunity, most notably being comparable to the waning of vaccine-induced immunity which has observed in many human diseases (Heffernan and Keeling, 2009) such as measles (Mossong et al., 1999), pertussis (van Boven et al., 2000), malaria (Okosun et al., 2011), and varicella (Chaves et al., 2007) and modeled using different methods. Given the broader applicability of our methods, our results have important implications for disease management, as we identify the most impactful parameters on disease outbreaks, and thus crucial intervention points.

Our sensitivity analysis found that the most influential parameters on both initial outbreak amplitude and time to initial outbreak in our system were parasite increase rate ($\mu$) and maximum increase rate of guppy immunity ($\theta$). These results indicate parasite growth rate and host resistance play the strongest role in the severity and speed of an outbreak, which makes logical sense. The higher parasite growth rate, or lower the immune response, the greater the parasite abundance will be. Given the small population sizes of our experiments, it is possible that host density may have affected the relative importance of these variables compared to the transmission rate or population size. The density of fish in our experiments was higher than wild populations (Croft et al., 2003), but lower than in commercial guppy populations (Kaiser et al., 1998), therefore an average approximation of the different conditions in which *Gyrodactylus* outbreaks may occur. Given the relatively short timescale of our experiments, we did not observe any impacts of longer-term parameters such as guppy birthrate or natural guppy mortality. However, sensitivity analysis of our generic outbreak magnitudes and periods consistently demonstrate that parasite virulence and parameters relating to guppy immunity have the strongest impact on our system and therefore this result was not an artifact of our experimental design. Both guppy resistance (Van Oosterhout et al., 2003; Dargent et al., 2016), and parasite virulence (Cable and van Oosterhout, 2007) are known to evolve rapidly and vary widely among populations due to different selective pressures and our findings indicate that understanding this heterogeneity is of significance to predicting and controlling disease outbreaks.

One limitation of this model is that it was based on laboratory, rather than field data and large differences in both host mortality and parasite burdens have been observed between the lab and field settings. Gyrodactylids persist in the wild and are observed at typically low burdens are observed, however mark-recapture experiments have suggested that infection can cause severe mortality (Van Oosterhout et al., 2007) typical of epidemics and in aquaculture (Johnsen, 1978; Johnsen and Jenser, 1991), and laboratory (Scott and Anderson, 1984; Van Oosterhout et al., 2003) settings, outbreaks are known to cause severe disease and mortality. Also, in the wild, guppies inhabit streams which are punctuated by "pools" separated by waterfalls, thus creating a network of populations among which unidirectional migration of hosts (and potentially parasites) downstream is possible (Van Oosterhout et al., 2007; Barson et al., 2009), however in our current model we focus only on populations in isolation. Furthermore, as previously mentioned, the population sizes were much smaller than those in a natural setting, and as such, the timescales used to estimate some of the longer-term parameters of our model, such as host birth and immunity waning to full susceptibility may not fully reflect dynamics in the wild (Van Oosterhout et al., 2007). Nevertheless, our estimates obtained from both literature and short-term data show a good fit for our data and could potentially be directly applied to aquaculture settings with only reparameterization specific to the species of interest. Moreover, longer-term simulations with our model show its ability to predict longer-term fluctuating dynamics such as those observed in the wild, indicating the predictive value of this model to more natural settings.

**TABLE 5** | The sensitivity of a generic outbreak period to the parameters.

| Parameter | Definition | Sensitivity index |
|---|---|---|
| $\beta$ | Transmission rate | −0.2360 |
| $\lambda$ | Recovery rate | −0.2712 |
| $\mu$ | Parasite increase rate | −0.5972 |
| $r$ | Half-saturation constant of per-capita parasite killing rate | 0.1035 |
| $\gamma$ | Maximum parasite killing rate | −0.1115 |
| $\theta$ | Maximum rate of increase of immunity | 0.5502 |
| $\kappa$ | Half-saturation constant of increase of immunity | 0.1037 |
| $\nu$ | Rate of decay of immunity in the absence of parasites | −0.3536 |
| $\alpha$ | Birth rate | −0.1824 |
| $d$ | Natural guppy mortality | 0.1708 |
| $K$ | Half-saturation constant for guppy growth | 0.2089 |
| $c$ | $1/c$ is the average duration of immune memory | 0.08776 |
| $\varepsilon$ | Parasite-induced mortality rate | −0.1128 |
| $\omega$ | Natural parasite mortality | −0.0537 |

Another limitation is our assumption of homogeneity of hosts, which is not accurate in this system. Guppies are known to exhibit a broad range in both life history traits (Reznick and Endler, 1982; Gordon et al., 2009) and innate resistance to parasites (Fraser et al., 2009; Fraser and Neff, 2010; Dargent et al., 2013), both within and among populations. Additionally, individual guppies may vary in their susceptibility to parasites due to individual characteristics such as size (Cable and van Oosterhout, 2007; Tadiri et al., 2013), carotenoid coloration (Grether et al., 2004; Kolluru et al., 2006) and sex (Richards et al., 2010, 2012; Dargent et al., 2016; Tadiri et al., 2016). Our parameters don't capture the wide variability that occurs in nature, or how this heterogeneity may influence host-parasite dynamics in the population but were instead based on average values obtained from literature and our own laboratory observations. Moreover, significant variability even in average population-level resistance has been observed among wild populations and domestic fish to various strains of gyrodactylids (Van Oosterhout et al., 2003; Cable and van Oosterhout, 2007; Dargent et al., 2013, Pérez-Jvostov et al., 2015) and it's possible that our estimated parameters may not fit some extreme cases of particularly low- or high-resistance population. However, despite not accounting for such complexities, our model fit data from two experiments, one which used fish from various wild populations from Trinidad and one which used domestic fish, therefore we find these average values to be a decent approximation.

In conclusion, we were able to develop and validate a mathematical model that more effectively describes the guppy-*Gyrodactylus* system, thus contributing to a further understanding of disease dynamics. Through sensitivity analysis, we were able to identify key factors affecting outbreaks to strategize control measures for parasites which increase in numbers due to reproduction directly on the host (in the absence of transmission) and are directly transmitted via host contact, particularly those relating to parasite growth rate and host resistance. Our findings have implications for a broader range of systems, with our model being most directly applicable to other gyrodactylids such as *G. salaris*, which is known to cause severe mortality and morbidity in Atlantic salmon fisheries (Johnsen, 1978; Bakke et al., 2007), but these methods could be also applicable to many other infections for which immunity decays over time, such as that observed for some vaccines.

## DATA AVAILABILITY

All datasets generated for this study are included in the manuscript and/or the **Supplementary Files**.

## ETHICS STATEMENT

The animal study was reviewed and approved by McGill University Animal Care Committee (AUP 2014-7547) in compliance with the Canadian Council on Animal Care.

## AUTHOR CONTRIBUTIONS

The mathematical model was developed at a meeting between CT, JK, GF, and HW. CT collected the data (either experimentally or from published literature) used for parameter estimates and model fitting, while JK validated the model and conducted the sensitivity analysis. CT wrote the initial draft of the manuscript, with editing input from all co-authors.

## FUNDING

We acknowledge support from FQRNT Equipe grant (MS and GF), NSERC Discovery grants (MS, GF, and HW), NSERC Postdoctoral Fellowship (JK) and NSERC Alexander Graham Bell Canada Graduate Scholarship (CT) and the McGill Department of Biology which provided travel funding for meetings between co-authors.

## ACKNOWLEDGMENTS

We thank DIMACS for providing space to conduct the analyses (partially enabled through support from the National Science Foundation under grant #CCF-1445755) as well as Simon Levin's Lab-Princeton University for the same purpose.

## SUPPLEMENTARY MATERIAL

The Supplementary Material for this article can be found online at: https://www.frontiersin.org/articles/10.3389/fevo.2019.00307/full#supplementary-material